\documentclass[journal abbreviation]{copernicus}
\usepackage{graphicx}

\begin{document}

\title{Homochirality through Photon-Induced
Melting of RNA/DNA: the Thermodynamic
Dissipation Theory of the Origin of Life}
\author{K.~Michaelian}
\maketitle

\begin{abstract}The homochirality of the
molecules of life has been a vexing problem
with no generally accepted solution to date.
Since a racemic mixture of chiral nucleotides
frustrates the extension and replication of
RNA and DNA, understanding the origin of
homochirality has important implications to
the investigation of the origin of life.
Theories on the origin of life have generally
elected to presume an abiotic mechanism
giving rise to a large prebiotic enantiomer
enrichment. Although a number of such
mechanism have been suggested, none has
enjoyed sufficient plausibility or relevance
to be generally accepted. Here we suggest a
novel solution to the homochirality problem
based on a recently proposed thermodynamic
dissipation theory for the origin of life. 
The ultraviolet absorption and
dissipation characteristics of RNA/DNA point
to their origin as photoautorophs,
their replication assisted by UV light and temperature, and acting as
catalysts for the global water cycle.
Homochirality is suggested to have been
incorporated gradually into the emerging life
as a result of asymmetric right- over
left-handed photon-induced denaturation of
RNA/DNA occurring when Archean sea surface
temperatures became close to the denaturing
temperatures of RNA/DNA. This differential
denaturing success would have been promoted
by the somewhat right-handed circularly
polarized submarine light of the late
afternoon when surface water temperatures are
highest, and a negative circular dichroism
band extending from 220 nm up to 260 nm for
small segments of RNA/DNA. A numerical model
is presented demonstrating the efficacy of
such a mechanism in procuring 100\%
homochirality of RNA or DNA from an original
racemic solution in less than 500 Archean
years assuming a photon absorption threshold
for replication representing the hydrogen
bonding energies between complimentary strands. 
Because cholesteric
D-nucleic acids have greater affinity for
L-amino acids due to a positive structural
complementarity, and because
D-RNA/DNA+L-amino acid complexes also have a
negative circular dichroism band between 200
- 300 nm, the homochirality of amino acids
can also be explained by the theory.
\end{abstract}

\sloppy
\affil{Instituto de F\'{\i}sica, Universidad
Nacional Aut\'{o}noma de
M\'{e}xico, Cto.~de la Investigaci\'{o}n
Cient\'{\i}fica, Cuidad
Universitaria, Mexico}

\correspondence{K.~Michaelian
(karo@fisica.unam.mx)}

\runningtitle{Homochirality through photon-induced melting of RNA/DNA}
\runningauthor{K.~Michaelian}

\received{9 February 2010} \accepted{21
February 2010} \published{}

\firstpage{1}

\introduction
Molecules that have no plane of symmetry come
in two distinct geometrical, but energy
degenerate, forms, or mirror images, called
``enantiomers", which are labeled as being
left (L)- or right (D)-handed. This chirality
in biological molecules is a result of the
tetravalent nature of carbon atoms, often
associated with a so-called ``alpha-carbon
atom" that attaches to a functional group.
Energy degeneracy implies that
enantiomers have essentially equal formation
and degradation probability under near
equilibrium conditions (except perhaps for
one part in 10$^{17}$, due to the parity
non-conserving weak force). However, life has
an overwhelming preference for one enantiomer
over the other, and thus non-equilibrium
biochemical reactions are chirality biased.
For example, RNA, DNA, ribose, and
deoxyribose are right-handed, while the amino
acids of life are left-handed.

Today, incorporation of only the correct
enantiomer of the nucelotides into RNA/DNA is
guaranteed by an unfailing chiral enzymatic
selection process. Such enzymes, however,
could not have been available at the very
beginnings of life. Without enzyme selection,
RNA template extension is severely adversely
affected in a racemic (equal concentration of
both enantiomers) solution of nucleotides,
principally because incorporated nucelotides
of the wrong chirality act as extension
terminators (Joyce et al., 1984). Orgel
(2004), has suggested that this frustration
during the copying of polynucleotides is one
of the greatest obstacles to an understanding
of the origin of life.

The simplest potential solution to the puzzle
of biotic homochirality has been to suggest
an initial overwhelming predominance of one
enantiomer over the other in the original
prebiotic soup, or, perhaps, a smaller
initial enantiomer excess subsequently
amplified by asymmetric autocatalysis
(Shibata et al., 1998) or other similar
mechanisms. Although many mechanisms for such
an original bias and amplification have been
proposed, none has come to be generally
accepted for lack of demonstrated relevance
or plausibility.

In the following section we briefly review
the mechanisms hitherto proposed for
enantiomer enrichment. Equally briefly, we
describe the problems associated with the
efficacy of each mechanism. In section 3 we
describe how homochirality could have arisen
gradually within the first replicating
organisms due to an asymmetry in the UV
photon-induced denaturation of RNA or DNA
(Michaelian, 2010), without the need to
invoke a prebiotic enantiomer excess or
catalytic amplification mechanism. Section 4
presents a simple model demonstrating the
efficacy of the proposed mechanism. Section 5
reviews the evidence for how D-nucleic acid
could have selected for L-amino acids.

\section{Prevalent Homochirality Theories}
Many theories for homochirality have proposed
a prebiotic enantiomer excess of the
biological molecules, either generated at the
Earth's surface or in space. Potential
mechanisms for generating this excess are;
circularly polarized light either
photolysing, photocatalysing, or
photoreacting the molecules; inorganic chiral
clay or crystalline template selectivity;
magnetochirality; and the parity violating
weak interaction.

Terrestrial circularly polarized light can be
generated by distinct mechanisms (Wolstencroft, 2004, Jorissen
and Cerf, 2002).  In order of importance these are: 1) Sunlight scattered at 
depth in water becomes linearly polarized. If this light is then totally
internally reflected at the water-air interface, its vertical
component undergoes a phase shift. As observed from below, near the surface,
one sees partially circularly polarized light outside Snell's window of up 
to 10\% (Wolstencroft, 2002; Horvath et
al., 2003). 2) Molecular (Rayleigh) atmospheric scattering gives linearly polarized
light and a subsequent aerosol (Mie) scattering ($r>>\lambda$) gives circular polarization. 
Right-handed circularly polarized light has
been measured at twilight up to a maximum of about 
0.5~\% (Angel et al., 1972; Wolstencroft, 1985).
3) Sunlight interacting with the Earth's magnetic field gives circular polarized
light through the Faraday effect. 3) The intrinsic circular polarization of
sunlight itself, about one part in $10^6$ (Kemp et al., 1987).

The maximum optical purity (enantiomer
excess) that could be obtained through
photocatalysing or photoreacting is given by
$g/2$ with $g=\Delta e/e$ where $\Delta e =
e_R - e_L$ and $e_R$ and $e_L$ are the molar
absorption coefficients for right and left
circularly polarized light, and $e=(e_R +
e_L)/2$ is the average of these. Empirical
studies suggest that the values of $g/2$ for
many different reactions are such as to
result in optical purity of usually less than
1~\% (Bonner and Rubenstein, 1987). It is
improbable, therefore, that circularly
polarized light could have given rise to
homochirality on Earth through
photocatalysing or photoreactions without
some kind of post amplification.

On the other hand, optical purity is not
limited to $g/2$ for photolysing, and thus
this has been the mechanism most studied.
However, there are three basic difficulties
with enantiomer enhancement on Earth through
photolysing. First, because the circular
polarization of sunlight is small, and since
the differential left- or right-handed
photolysing capacity is small, a very large
amount of basic RNA/DNA material would have
to be destroyed in order to obtain 100\%
chirality. Experiments with camphor, for
example, suggest that 20\% chirality can be
achieved by photolysing 99\% of the original
racemic material (Balavoine et al., 1974).
Homochirality would thus require essentially
complete destruction of the original
material. Secondly, averaged over the full
diurnal cycle, the net circular polarization
of sunlight is zero. Finally, high
temperature, metal ions, radiation, and
ultraviolet light itself, all have the
tendency to cause racemization, and this
effect is enhanced if the molecules are in
water (Schroeder and Bada, 1976).

The smallness of the terrestrial circular
polarization of sunlight, and its averaging
to zero over the diurnal cycle, has led
investigations into considering an
extraterrestrial origin of the basic
molecules of life and their chirality.
Astronomical sources of potentially much
greater circular polarization and intensity
have been proposed, such as synchrotron
radiation from neutron stars with large
magnetic fields (Bonner and Rubenstein, 1987;
Cline, 2005). However, very few such circularly
polarized light sources have been found to date,
and all, so far, have only been identified in
the infrared (albeit, presumably because
shorter wavelength light does not penetrate
as well the extensive dust clouds of space).
Furthermore, the synchrotron radiation form
these sources is generally white, and
non-trivial frequency dependent dispersion
properties of the organic molecules means
that circular dichroism (differential
absorption of left- over right-handed
circularly polarized light) is both positive
and negative in different regions of the
spectrum. In fact, integrated over the whole
spectrum circular dichroism sums to zero,
the ``Kuhn-Condon zero sum rule" (see section
3 and Mason, 1988). Therefore, a net
enhancement of either chirality could only be
entertained if additional arguments
restricting the extraterrestrial light to a
relevant frequency range could be found
(Bonner et al., 1998).

Finally, it is known that gamma rays and high
energy particles, as well as unpolarized UV
light itself, cause radioracemization
(equalization) of the original enantiomer
excess (Keszthelyi, 1995, Cataldo et al.,
2005), so further mechanisms would have to be
identified which could keep the molecules in
their chiral state during their trip to
Earth. Notwithstanding these difficulties,
however, up to 15~\% enantiomer excess has
been claimed for some non-biological amino
acids delivered to the Earth in carbonaceous
chondrite meteorites such as Murchinson.
Biological amino acids found in these
meteorites have little, if any, enantiomer
excess (Pizzarello et al. 2003).

It has been suggested that inorganic elements
crystallizing with a preferred chirality could
have acted as templates for generating the
chirality bias of the molecules of life.
Bonner et al. (1974, 1976) found that
amino acids are enantioselectively adsorbed
on chiral,
enantiopure quartz crystals. For example,
D-Alanine is bound selectively
to D-quartz with an enantiomer excess of up
to 20\%. Results
of several groups claiming to have found a
selective adsorption of
amino acids on the surfaces of achiral clays
have been controversial (Bonner et al., 1974,
1976) . Although there is evidence of a very small
chiral selectivity by clay minerals, it has been argued that such a
small effect may be due to previous
absorption of optically active biomolecules
produced by living organisms (Youatt and
Brown, 1981). It is still uncertain, but
unlikely, that prebiotic clays could have had
a chiral bias.

Illumination of a racemic mixture of chiral
molecules in a magnetic field by
non-polarized light induces an enantiomer
excess through the Faraday effect (Rikken and
Raupach, 2000; Barron, 2000). This so caller
``magneto-chiral dichroism" is operative on
Earth, generating circularly polarized light
from the interaction of unpolarized sunlight
with the terrestrial magnetic field. However
the anisotropy factor is small, of order
$10^{-10}$ (Jorissen and Cerf, 2002). A
further problem is that the magneto-chiral
dichroism effect has opposite sign on
opposite sides of the equator. Very young stars have a large
magnetic field due to high rotation rates and
are also sources of intense UV light. A
magneto-chiral effect produced by such a star
would be larger than that due to a
terrestrial source, but still small, giving
rise to an enantiomer excess of only about
$10^{-6}$ (Jorissen and Cerf, 2002).

The weak force is parity violating, resulting
in a breaking of the energy degeneracy of the
right- and left-handed enantiomers, thus
favoring one over the other. This was first
proposed to be the source of biomolecular
homochirality by Ulbricht (1957). However, a
comparison of the weak energy to thermal
energy at the Earth's surface gives $\Delta
E/k_BT\approx 10^{-17}$ (Cline, 2005), much
too small to be a plausible solution in
itself to homochirality. Vester et al. (1958)
proposed a somewhat different mechanism for
an enantioselective reaction originating from
the parity violating weak interaction.
According to the ``Vester-Ulbricht
hypothesis´´, the longitudinally polarized
$\beta$-decay electrons would, when
decelerated in matter, lead to circularly
polarized bremsstrahlung photons, promoting
enantioselective reactions. However, as
mentioned previously, enantiomer excess is
limited to $g/2$,
which, for most relevant reactions, is very small.

In summary, although many mechanisms could
have given rise to a small enantiomer excess locally,
and during a finite time period, these
alone would not have been sufficient to lead
to the homochirality of life. An additional
auto-catalytic amplification mechanism
(Shibata et al., 1998), or far from
equilibrium condition (Kondepudi, 1987,
Micheau et al., 1987) would have been needed
to bring the effect to the level of
homochirality. Amplification mechanisms rely
on different barrier heights in chemical
reactions involving chiral catalysts of a
small enantiomer excess. However, in true
thermodynamic equilibrium, the products must
necessarily be racemic, independently of
barrier heights, but if the reaction is
incomplete, or driven out of equilibrium,
then one of the product enantiomers could be
produced, at least in the short term, in much
greater quantity than the other (Podlech,
2001). Far from equilibrium theories rely on
spontaneous symmetry breaking, a type of
second order phase transition involving a
control parameter which passes through a
critical value. Spontaneous symmetry breaking
through amplification of a microscopic
fluctuation in non-equilibrium systems with
non-linear kinetic laws has been demonstrated
by Prigogine (1967).

Amplification, by whatever mechanism,
therefore requires a non-equilibrium
situation. Indeed, since life is an out of
equilibrium phenomena, it is not surprising
that many of life's enzymatic promoted
chemical reactions are chirality biased.
Although such ideas for homochirality have been
argued to apply in general, and although there exists
experimental evidence validating the idea for
certain non-equilibrium chemical reactions
(see Podlech, 2001 and references therein),
there has as yet been no demonstration of the
principle in association with the putative
original molecules of life;
the amino acids or the nucleic acids RNA and
DNA.

\section{Homochirality through photon-induced
melting of RNA and DNA}
   
The Earth's surface during the Archean
(3.8-2.5 Ga) was subjected to intense
ultraviolet light within the 200-300 nm
wavelength region (Sagan, 1973, Cnossen et
al., 2007), the result of a young Sun
(Tehrany et al., 2002) and to the lack of UV
absorbing oxygen and ozone in the Earth's
atmosphere. RNA and DNA are extraordinary
absorbers and dissipators of UV light within
this spectral region (Middleton et al.,
2009). According to the thermodynamic
dissipation theory of the origin of life
(Michaelian, 2010), life arose as a catalyst
for the water cycle by absorbing this light and transforming it into heat,
thereby augmenting the daytime temperature of
the ocean surface. Circumstantial evidence
exists indicating that RNA and DNA were
photoautotrophs, obtaining their free energy
for assembly and reproduction from the
intense ultraviolet light, while at the same
time, by coupling to the water cycle,
producing much more entropy than attributable
to their metabolism and replication
(Michaelian, 2010). Such a view connects the
visible light dissipation by plants and
cyanobacteria today with UV light dissipating
RNA/DNA in the Archean, emphasizing life's
continued involvement in the water cycle.

Geochemical evidence in the form of
$^{18}$O/$^{16}$O ratios found in cherts of
the Barberton greenstone belt of South Africa
point to an Earth's surface temperature of
around $70\pm 15$\thinspace $^{\circ }$C
during the 3.5--3.2\thinspace Ga era (Lowe
and Tice, 2004). These temperatures, near the
beginnings of life (ca. 3.8 Ga), are close to
the melting temperatures of RNA and DNA. An
enzyme-free mechanism for replication can
therefore be imagined in which absorption and
dissipation of UV light into heat by the
nucleic acids during the day increased the
local sea-surface temperatures to beyond the
denaturing temperature of RNA or DNA,
allowing the separated strands to act as
templates during the cooler periods overnight
(Michaelian, 2010). Such ultraviolet and
temperature assisted replication (UVTAR)
bears similarity to polymerase chain reaction
which is used to amplify particular segments
of RNA or DNA in the laboratory (Mullis,
1990). Photon-induced RNA/DNA denaturation
has been experimentally detected (Hagen et
al., 1965, Roth and London, 1977).

Replication of RNA and DNA could therefore have been
promoted by the local diurnal variation of the
sea surface temperature, due in large part to
the absorption and dissipation of UV light by
RNA and DNA at the ocean surface. Enzymes,
and thus information content and reproductive
fidelity, were not required until the sea
surface temperature had cooled to somewhat
below the melting temperature of RNA and DNA.
Longer RNA or DNA segments that later coded
for simple denaturing enzymes could continue
replicating at colder temperatures, thereby
initiating evolution through natural
selection in response to a cooling ocean
surface (Michaelian, 2010).

Scattering of unpolarized UV
sunlight from water molecules and suspended 
particles and a subsequent total internal reflection
of this light at the air-water interface, would have 
led to a component of about 5 \% right-handed 
circular polarization during the afternoon near the 
surface (Wolstencroft, 2004),
independently of the hemisphere, season, or
terrestrial magnetic polarization reversals.
Since the sea surface temperature would be
greatest in the late afternoon, this fact
could have contributed to an enhancement of
RNA/DNA with D-enantiomer nucleotides because
of the unequal absorption cross sections for
left- and right-handed circularly polarized
light on these chiral molecules. Double
strands containing L-enantiomer nucleotides
would have been at a disadvantage since they
would absorb less well the right-handed
circularly polarized light of the late
afternoon, and thus could not raise local
water temperature as often for denaturation.
These, therefore, would suffer from a
somewhat lower probability of reproduction
through UV and temperature assisted
replication. Once RNA/DNA containing
predominantly L-enantiomer nucleotides had
formed, they would tend to become locked in
the double strand formation, effectively
removing them as templates for facilitating
further reproduction. Those with mainly
D-enantiomer nucleotides could have continued
replicating, and thus evolving.

The nomenclature, ``left-´´ or
``right-handed´´, is often not related to the true
optical chirality of molecules. Optical
chirality is collective to many asymmetric
centers while the nomenclature refers to
usually only one of them. Therefore, a
quantitative theory of homochirality, based
on differential absorption of circularly
polarized light, requires a careful look at
the full circular dichroism (CD) spectrum as
a function of wavelength for RNA and DNA and
their complexing with amino acids.

The CD spectrum of DNA and RNA depends on
temperature, salinity, and pH. Higher
temperature has the effect of reducing the
amplitude of the circular dichroism with
little effect on peak position or zero
crossings (Gray et al., 1978). At neutral pH,
the CD spectrum of DNA shows a negative band
(greater absorption of right-circularly
polarized light) with a maximum at 245 nm,
extending to about 260 nm, and a positive
band with a maximum at approximately 275 nm
(Hillen et al., 1981). The negative band has
been shown to be a result of base stacking
(Sprecher et al., 1979 ) and is relatively
independent of base content and secondary
structure, while the positive band depends on
these characteristics (Hillen et al., 1981).
The CD spectrum of shorter polynucleotides
shows a wider negative CD band spanning the
region of 220 to 260 nm at neutral pH (Gray
et al., 1978). It is thus probable that this
negative CD band was responsible for the
gradual accumulation of homochirality in
RNA/DNA through the ultraviolet and
temperature assisted mechanism described
above.

\section{Model Simulations}
A racemic mixture of single strand RNA/DNA
segments produced by UV photochemical
reactions on atmospheric gases probably
floated on the surface of a hot prebiotic
Archean ocean (see Michaelian, 2010 and
references therein). Assuming that these
segments could begin to act as templates for
reproduction when the sea surface
temperature at night dropped below their
melting temperature, we can estimate how
rapidly chirality would have grown in the population due to a
slightly greater absorption probability of
right- over left-handed circularly polarized
light at about 255 nm where RNA/DNA absorb 
most strongly. The model is made as simple as
possible, avoiding all unnecessary details.

For a fixed ocean
surface temperature, melting of RNA/DNA, and
therefore the possibility for replication, would be, in first
approximation, proportional to the amount of UV light
absorbed. In reality, an energy threshold
exists due to the hydrogen bonding between
strands. (This threshold has important consequences and 
will be considered in
an extension of the model given below.) The
following recursion relations then give the
number of left-handed and right-handed strands
($nL_i$ and $nR_i$) at any given diurnal cycle
$i$
\begin{eqnarray}
nL_i = nL_{i-1}(1+c(P_{LL} + P_{LR})),
\nonumber\\
nR_i = nR_{i-1}(1+c(P_{RR} + P_{RL})),
\label{eq:recurs} 
\end{eqnarray}
where $P_{LL}$ and $P_{LR}$ are the average
(over all existing strands) relative
probabilities that a left-handed RNA/DNA
absorbs a left- and right-handed photon
respectively. $P_{RR}$ and $P_{RL}$ are
similarly defined, but for absorption on a
right-handed RNA/DNA. All
probabilities are taken relative to that of
right-handed circularly polarized light
absorption on right-handed DNA, $P_{RR}$, being this
the largest since it may be assumed that most
melting would occur in the afternoon when surface water
is at highest temperature, and
afternoon light is somewhat right-handed circularly
polarized. Also, D-RNA/DNA have a negative
circular dichroism band between 220 and 260
nm where RNA/DNA absorbs strongly. Therefore,
\begin{eqnarray}
P_{RR}& =&1.0,\nonumber\\
P_{LL} &= &{(1.0-\Delta_{RCPL})\over
(1.0+\Delta_{RCPL})},\nonumber\\
P_{LR} &= &{(1.0 - \Delta_{RCD})\over
(1.0+\Delta_{RCD})},\nonumber\\
P_{RL} &=
&{(1.0-\Delta_{RCPL})\over(1.0+\Delta_{RCPL})}\cdot{(1.0
-\Delta_{RCD})\over(1.0+\Delta_{RCD})}.
\label{eq:prob} 
\end{eqnarray}
where $\Delta_{RCPL}$ is the right-handed circular
polarized light excess of the afternoon, and
$\Delta_{RCD}$ is the right-handed circular dichroism
excess. The $\Delta_{RCPL}$ of visible light in the afternoon today 
is only about 0.5\% (Angel et al.,
1972, Deutsch, 1991). However, that due to multiple scattering in water 
and totally internally reflected at the water-air 
interface is much greater, about 5\% (Wolstencroft, 2004).
According to Gray et al. (1978) the
differential absorption of right- over
left-handed photons $\Delta_{RCD}$ due to circular
dichroism at 250 nm for short polynucleotides
is about 4/6000. The $c$ in equation
(\ref{eq:recurs}) is a  normalization constant that would
depend on the daytime intensity of the
incident UV light, the concentration of
nucleotides available at the sea-surface at
night, the length of the RNA/DNA template
strand, the sea-surface temperature at night,
duration of night, etc. Until such values for
the Archean are better constrained, an exact
calculation cannot be made. However, for the
sake of argument, we take as a plausible
value of one in 1,000 RNA/DNA segments
reproducing through the UVTAR mechanism during each diurnal
cycle, i.e. $c=.001$ (in polymerase chain
reaction with an unlimited supply of nucleotides, 
primers, and the enzyme polymerase,
but much shorter extension times,
this value is very close to one).

\begin{figure}
\includegraphics[width=300px,height=260px]{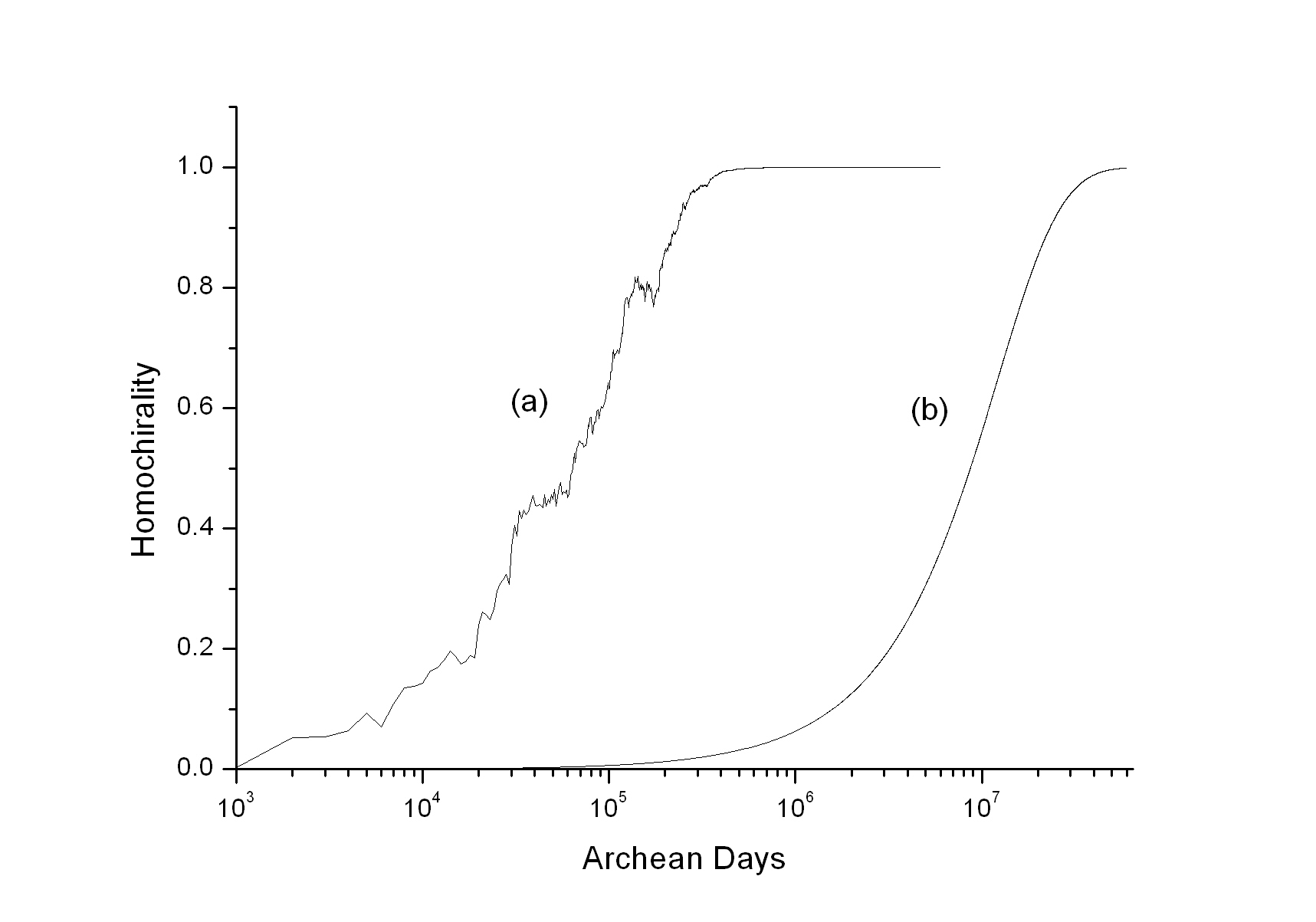}
\caption{Homochirality as a function of number of Archean days. (a) Including 
an energy threshold for denaturation related to the complimentary strand binding
energy, (b) assuming denaturation probability is simply proportional to
the number of photons absorbed.}
\label{fig:hc}
\end{figure}

Using these values in the recursion equation
(\ref{eq:recurs}) together with an equation
for the homochirality
as a function of diurnal cycle $i$ 
\begin{equation}
HC_i= - {nL_i - nR_i \over nL_i+nR_i},
\label{eq:hc} 
\end{equation}
gives curve (b) plotted in figure
\ref{fig:hc}. Given racemic initial values,
$nL_0$ and $nR_0$, the curve is independent
of these initial values.
Since an Archean day was about 1/2 the length
of an actual day, and assuming the orbit of
Earth has not changed, figure \ref{fig:hc}
implies that practically 100\% homochirality
can be obtained in less than 50,000 Archean
years.

The model can be refined by including an
energy threshold for denaturation. Such a
cut-off exists due to the specific temperature dependent
hydrogen bonding energies between the two
complimentary strands. The threshold would be larger in
the morning than afternoon because of a
cooler surface temperature. Denaturation of
right-handed RNA/DNA would therefore be
further favored by this threshold since there is more
right-handed circularly polarized light
available in the afternoon. This threshold can be
included in the model by randomly varying the
$P_{RR}, P_{LL}$, etc. somewhat about their
nominal values (to mimic the statistical
fluctuations of photon absorption) and
setting the combined probabilities for
denaturation $P_{LL} + P_{LR}$ and $P_{RR} +
P_{RL}$ to zero if they fall below a
specified limit. Such a threshold
dramatically increases the rate of obtainment
of homochirality, giving times as short as
500 Archean years (curve (a) of figure
\ref{fig:hc}). A 1\% replication rate, $c=0.01$, leads to
obtainment of homochirality in only 50 Archean years (not shown). A
greater right-handed circular polarization of submarine UV light during the
afternoon, than that assumed of 5\%, would decrease this time further.

\section{Homochirality of the amino acids}
If homochirality had been obtained for all
the 20 amino acids of life before
incorporation into the first replicating
organism, through, for example, photolysing
by circularly polarized light in space, then
a spectral window for the circularly
polarized UV light in which the circular
dichroism band was large and of the same sign
for all the amino-acids would have to exist.
An analysis of the CD data of amino acids by
Cerf and Jorissen (2000) suggests that such a
mechanism could not have been operating if
tryptophan or proline were among the original
amino acids. Furthermore, stability of the
amino acids against racemization would have
to be demonstrated. The $\alpha$-methyl amino
acids found with non-negligible chirality in
meteorites have sufficient stability against
racemization (Bada, 1991), but the
$\alpha$-hydrogen amino acids composing the
20 natural amino acids of today's proteins do
not (Pizzarello and Cronin, 2000).
Furthermore, experiments on photolysing of
the amino acids have demonstrated only weak
enantiomer excesses of a few percent (Bada,
1991 and references therein). Uncertainty
also remains as to whether life based amino
acids have yet been detected in space, where
photolysing possibilities may be greater
than on Earth.

A more plausible alternative for the
homochirality of the amino acids is chiral
discrimination by D-nucleic acids, resulting
from its structural complementarity with
L-amino acids. Evidence of chiral selectivity
of activated L-amino acids by DNA through
protein intercalation between adjacent base
pairs has been obtained by Barton et al.
(1982). Reich et al. (1996) have demonstrated
experimentally that D-nucleic acid in the
cholesteric form (with the molecule folded in
on itself) has greater affinity for the
poly-L-lysine than for poly-D-lysine. Also,
using molecular modeling techniques, Bailey
(1998) has shown that D-RNA constrained to a
surface selects preferentially for L-amino
acids. Such post chiral selectivity for the
amino acids has resonance with the
thermodynamic dissipation theory of the
origin of life because in this theory enzymes are postulated
to have arisen later in life's history as the
Earth's surface temperature cooled and
conditions began to stray from those
favorable to UV and temperature assisted
replication of RNA/DNA.

As sea surface temperatures cooled and its
salinity increased, longer RNA/DNA segments
would spontaneously take on cholesteric forms
(Reich et al., 1996) in which the
right-handed double-helix folds in on itself
to produce a supra-molecule with enhanced
right-handed asymmetry. The circular
dichroism of these cholesteric forms is
positive within the 200-300 nm region (Reich
et al., 1996) and these by themselves would
therefore not absorb as well the right-handed
circularly polarized light of the afternoon.
However, Reich et al. (1996) have also shown
that L-amino acids have a significantly
larger affinity to D-DNA in the cholesteric
form than do R-amino acids, and thus L-amino
acids would have been naturally selected by
these. These D-DNA+L-amino acid complexes
have, in fact, a negative circular dichroism over the
whole 200 to 300 nm region (Reich et al.,
1996), implying greater absorption
efficiency for the right-handed circularly
polarized light. These D-DNA+L-amino acid
complexes would then be more able at raising
the local water temperature to beyond the
denaturing temperature, providing the
templates for reproduction during
cooler periods.

\section{Conclusions}
Hitherto proposed mechanisms for the
homochirality of the biomolecules, mostly invoking an
abiotic mechanism for producing a
prebiotic enantiomer enrichment, have not
been generally accepted for lack of
plausibility or relevance. The thermodynamic
dissipation theory for the origin of life
offers a novel possibility in which the mechanism for the obtainment of
homochirality is incorporated into the replication
mechanism for emerging life through asymmetric right-
over left-handed photon-induced denaturation
of RNA/DNA due to a negative circular
dichroism band extending from 220 nm up to
260 nm for small segments.
Photon-induced denaturation would be much more
effective in producing homochirality than photoreaction,
photocatalysing, or photolysing because it
deals with weak hydrogen bonds rather than
strong covalent bonds, and further, by
operating close to the denaturing temperature
of RNA/DNA, there exists a temperature dependent
threshold related
to the strength of these bonds which becomes
greater as the sea surface cools,
favoring still more D-RNA/DNA. The mechanism,
in analogy with polymerase chain reaction, but unlike
previously proposed mechanisms, produces an
exponential increase in chirality in the population
with diurnal cycle. Homochirality, the ratio
of the difference of populations to their
sum, thus increases linearly while the enantiomer
populations are similar (Fig. \ref{fig:hc}). This is a case of a
far from equilibrium process operating under
varying boundary condition, rather than an
example of a non-equilibrium spontaneous
symmetry breaking process.

The most plausible scenario for the
homochirality of the amino acids is that of
chiral selectivity of D-nucleic acid for
L-amino acids due to complementarity of
structure, particularly when DNA is in its
folded cholesteric form, of relevance to
longer strand RNA/DNA. This, in turn would
have relevance to post origin of life colder sea surface
temperatures when enzymes to aid denaturation
became necessary (Michaelian, 2010). D-DNA+L-amino acid
complexes have negative circular dichroism
over the entire 200 to 300 nm region, while
D-DNA+R-amino acid complexes have positive
circular dichroism over this region (Reich et
al., 1996). D-DNA+L-amino acid complexes
would thus have greater replication
probability under the UV and temperature
assisted replication theory.

\begin{acknowledgements}
The financial assistance of DGAPA-UNAM, 
grant numbers~IN118206 and~IN112809 is
greatly appreciated.
\end{acknowledgements}

\end{document}